\documentclass[prl,twocolumn,showpacs,preprintnumbers,aps,nofootinbib]{revtex4-1}
\usepackage{graphicx}% Include figure files
\usepackage{dcolumn}% Align table columns on decimal point
\usepackage{bm}% bold math
\usepackage{amsmath,amssymb,amsfonts}
\usepackage{latexsym}
\usepackage{bbm}
\usepackage{url} 
\usepackage{color}
\usepackage{soul}
\usepackage{multirow}
\usepackage[caption=false]{subfig}
\usepackage{braket}
\usepackage{hyperref}

\begin{document}
\captionsetup[subfigure]{labelformat=empty}
\preprint{LA-UR-23-25748}

\title{ALP contributions to $\mu\to e$ conversion }

\author{Kaori Fuyuto}
\email{kfuyuto@lanl.gov}
\author{Emanuele Mereghetti}
\email{emereghetti@lanl.gov}
\affiliation{Theoretical Division, Los Alamos National Laboratory, Los Alamos, NM 87545, USA}

\bigskip

\date{\today}

\begin{abstract}
We study the $\mu\to e$ conversion process in nuclear targets arising in models of axion-like particles (ALPs) with hadronic and charged lepton flavor violating (CLFV) interactions. Contributions to this process generally fall into two categories: spin-independent (SI) and spin-dependent (SD). 
While the SI contribution can be generated by a dipole operator through purely leptonic ALP interactions, the SD contribution can also be present through ALP-quark interactions at tree-level. It is naively anticipated that the SI contribution would be dominant due to its coherent enhancement. In this {\it letter}, we show that is not generically the case; in particular, for naturally-sized ALP couplings to quarks of order $\sim\!m_q/f_a$, the SD interaction induced by ALP-$\pi^0$ mixing turns out to be the leading contribution to $\mu\to e$ conversion. Intuitively, this stems from the suppressed dipole contribution by the QED one-loop factor which counters the effect of SI coherent enhancement.
Our study highlights the importance of $\mu\to e$ conversion searches in exploring the parameter space of generic ALP models, and demonstrates the competitiveness of these searches in probing the CLFV ALP parameter space in the heavy mass range of $m_a\gtrsim m_\mu$.
\end{abstract}

\maketitle

%-----------------------------------------------------------------------------------------------------------------------------------
%	Introduction
%-----------------------------------------------------------------------------------------------------------------------------------
\paragraph*{Introduction---}
Low energy processes that are either forbidden or highly suppressed in the Standard Model (SM) can be powerful probes of physics beyond the Standard Model (BSM).
One example is charged lepton flavor violation (CLFV).
While in the SM with massive neutrinos charged-lepton-flavor  is violated, CLFV amplitudes are so severely suppressed by powers of $m_\nu/m_W$ \cite{Marciano:1977wx,Petcov:1976ff} that the observation of any such process would necessarily imply the existence of BSM. Currently, the most stringent limit on CLFV comes from bounds on $\mu\to e$ transition, ${\rm BR}(\mu^+\to e^+\gamma)<4.2\times 10^{-13}$ \cite{MEG:2016leq}.
This bound is expected to be improved by the MEG II experiment down to ${\rm BR}(\mu\to e \gamma)\sim O(10^{-14})$ \cite{MEGII:2018kmf, MEGII:2021fah}. Concurrently, $\mu\to e$ conversion searches at Fermilab (Mu2e) and J-PARC (COMET) aim at improving the current bound on ${\rm BR}(\mu\to e)$ by {\it four orders} of magnitude, down to ${\rm BR}(\mu\to e)\sim {\cal O}(10^{-17})$ \cite{Mu2e:2014fns, Mu2e:2018osu, COMET:2018auw}.

Several phenomenological studies have approached the $\mu \to e$ conversion process in the framework of Effective Field Theory (EFT) assuming that CLFV sources originate from ultraviolet physics above the electroweak scale \cite{Weinberg:1959zz, Shanker:1979ap, Czarnecki:1998iz, Kitano:2002mt, Cirigliano:2009bz, Rule:2021oxe, Cirigliano:2022ekw, Haxton:2022piv}. These studies have focused mainly on spin-independent (SI) contributions, which are coherently enhanced by $Z^2$, where $Z$ is the nucleus atomic number. The subdominant effect of spin-dependent (SD) interactions have been studied in \cite{Cirigliano:2017azj, Davidson:2017nrp, Crivellin:2017rmk, Davidson:2018kud, Davidson:2020hkf, Rule:2021oxe, Haxton:2022piv,Hoferichter:2022mna}.

More recently, CLFV originating from light new physics at the GeV scale or below has also been considered. A particularly well-motivated class of particles with CLFV are {\it axion-like particles}, or ALPs \cite{Calibbi:2016hwq, Calibbi:2020jvd, Linster:2018avp, Ema:2016ops, Froggatt:1978nt, Heeck:2019guh, Garcia-Cely:2017oco, Ibarra:2011xn}, which are naturally light since they originate from the spontaneous breaking of a global symmetry. In particular, a light enough ALP $a$ with $m_a\lesssim m_\mu-m_e$ can be emitted on-shell in rare muon decays, $\mu\to e a$. This signal is strongly constrained by bounds on the branching ratio ${\rm BR}(\mu\to e a)<O(10^{-6})$ \cite{Jodidio:1986mz, TWIST:2014ymv}, which translate into an upper bound on the ALP decay constant of $f_a\gtrsim\mathcal{O}(10^9)$~GeV \cite{Calibbi:2020jvd}.  The upcoming experiments MEG II and Mu3e are expected to reach ${\rm BR}\sim O(10^{-8})$ \cite{Calibbi:2020jvd, Perrevoort:2018okj}.

%---------------- Figures ---------------------
\begin{figure}[t]
\includegraphics[width=\columnwidth]{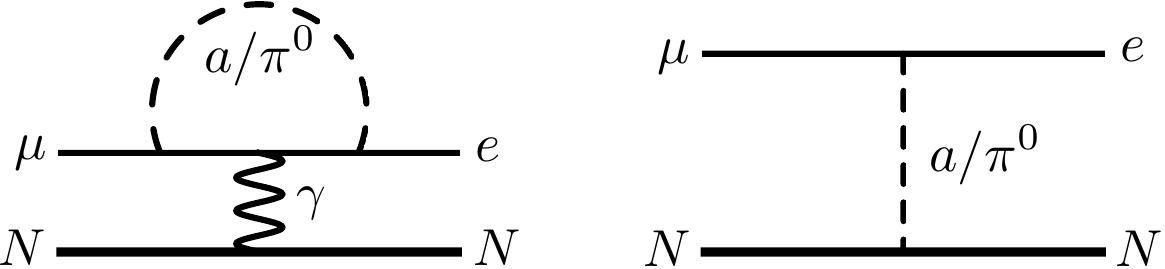}
\caption{Contributions to $\mu\to e$ conversion in nuclei from the LFV 1-loop dipole operator (left), and $t$-channel exchange of the ALP $a$ and $\pi^0$ (right).}
\vspace{-1em}
\label{dipole_4f}
\end{figure}
%------------------------------------------------

In contrast, for heavier ALPs with $m_a\gtrsim m_\mu-m_e$, $\mu\to e\gamma$ and $\mu+N\to e+N$ become more relevant processes to probe CLFV ALP interactions. Naturally, these processes are sensitive to different couplings in the ALP parameter space. Ref.\,\cite{Cornella:2019uxs} has considered in detail {\it leptonic} ALPs with both flavor-diagonal and flavor-violating couplings. Here, we turn our attention to ALPs with additional hadronic couplings and their mediation of $\mu+N\to e+N$ processes. Previous studies in the literature have estimated the rate for this process by assuming that the coherently enhanced contribution from the dipole operator induced at {\it one-loop} level (left in Fig.\,\ref{dipole_4f}) was the dominant one. In this {\it letter}, we consider additional contributions to the conversion process from flavor-diagonal ALP-quark couplings. Our main finding is that amplitudes involving $t$-channel exchange of $a$ and $\pi^0$ (right in Fig.\,\ref{dipole_4f}) can strongly dominate the rate for $\mu+N\to e+N$, despite the fact that these amplitudes are not coherently enhanced. 
The main intuition for this is two-fold: (i) the contributions from tree-level $a/\pi^0$ exchange benefit from the relatively large axial coupling to nucleons, $g_{A}=1.27$, and (ii) the contribution from the dipole interaction is suppressed by one electromagnetic loop $\sim \alpha_{\rm em}/4\pi$.  Our study highlights the importance of $\mu\to e$ conversion experiments in discriminating between different ALP models.

\vspace{0.5em}
%-----------------------------------------------------------------------------------------------------------------------------------
%	Mu to e conversion
%-----------------------------------------------------------------------------------------------------------------------------------
\noindent\emph{$\mu\to e$ conversion---}
The interactions of an ALP $a$ with decay constant $f_a$ can be quite general and include couplings to gauge bosons as well as couplings to fermions, that can be both flavor-conserving and flavor-violating:
\begin{align}\label{LaGeneral}
&{\cal L}^a\supset-\frac{a}{f_a}\bigg[\sum_{X=F,Z,W,G}\,c_XX\tilde{X}\nonumber\\
&\hspace{-0.1cm}+\sum_{f,i,j}\bar{f}_i\big[i(m_{f_j}-m_{f_i})\,v^f_{ij}+i(m_{f_j}+m_{f_i})\,a^f_{ij}\gamma_5\big]f_j\bigg]
\end{align}
where the sum over $f$ runs over all quarks and leptons, $i,j$ are family indices, and the matrices $v^f_{ij}$, $a^f_{ij}$ are Hermitian.

Due to the plethora of parameters in (\ref{LaGeneral}), the dependence of BSM processes on the most general ALP parameter space can be quite complicated, and the individual contributions of different ALP couplings can be easily obscured. Therefore, in order to facilitate the comparison between tree-level $a/\pi^0$-exchange vs dipole contributions to $\mu+N\to e+N$, we shall focus on a simplified model of an ALP that has only pseudoscalar isovector couplings to light quarks, a flavor-diagonal coupling to muons, and flavor-violating couplings to $\mu$ and $e$:
 \begin{subequations}
 \begin{eqnarray}
{\cal L}^a\supset-i\frac{a}{f_a}\left(m_u\,\bar{u}\,\gamma_5u-m_d\,\bar{d}\,\gamma_5d+2a_{\mu\mu}\,m_{\mu}\,\bar{\mu}\,\gamma_5 \mu\right)~~&\label{diagonal_ALP}\\
-i\frac{a}{f_a}\,m_{\mu}\left[\bar{e}\,(v_{e\mu}+a_{e\mu}\gamma_5 )\mu\right]~~&\label{LFV_ALP}
 \end{eqnarray}
\end{subequations}
All other couplings $v^f_{ij}$, $a^f_{ij}$ not explicitly featured in \eqref{diagonal_ALP} and \eqref{LFV_ALP}, as well as gauge boson couplings are set to zero.

The ALP coupling to light quarks in (\ref{diagonal_ALP}) generates effective ALP couplings to hadrons below the QCD confinement scale. The most important parameter controlling these couplings is the $a-\pi^0$ mixing, which can be straightforwardly obtained by mapping (\ref{diagonal_ALP}) into chiral perturbation theory ($\chi$PT) and diagonalizing the light meson mass matrix. The physical states $a_p$ and $\pi^0_p$ are given in terms of the original states $a$ and $\pi^0$ by:
\begin{subequations}
\begin{align}
 a_p~&=~c_\alpha a -s_\alpha \pi^0\\
 \pi^0_p~&=~s_\alpha a+c_\alpha \pi^0,
\end{align}
\end{subequations}
where $c_\alpha$ and $s_\alpha$ are short-handed notation for $\cos\alpha$ and $\sin\alpha$, respectively, and $\alpha$ is the $a-\pi^0$ mixing angle.
We can then re-express the CLFV and nucleon couplings of the low energy physical states as \footnote{Direct couplings of 
the  pseudoscalar density, and thus of the axion, to the nucleon arise at next-to-next-to-leading order in chiral perturbation theory \cite{Fettes:1998ud}, and therefore we neglect them.}
\footnote{We express the pion-nucleon coupling in a pseudoscalar form. At the order we are working, this is equivalent to the axial form dictated by chiral symmetry \cite{Bernard:1995dp}, and it allows for a more easy relation to the calculation of the nuclear response functions of Refs. \cite{Haxton:2022piv,Rule:2021oxe}  }:
\begin{eqnarray}\label{Lphys}
&&\!\!\!\!{\cal L}^{\rm phys}\supset-g_A\frac{m_N}{f_{\pi}}\left(c_{\alpha}\pi^0_p-s_{\alpha}a_p \right)\overline{N}i\gamma_5\tau^3N\\
&&\!\!\!\!-i\frac{m_\mu}{f_a}\big(c_{\alpha}a_p+s_{\alpha}\pi^0_p \big)\left[2a_{\mu\mu}\,\bar{\mu}\,\gamma_5\mu+\bar{e}\,(v_{e\mu}+a_{e\mu}\gamma_5) \mu\right].\nonumber
%\\\nonumber
\end{eqnarray}
Above, $N=(p,n)^T$ is the nucleon isospin doublet, $\tau^3$ is a Pauli matrix, $g_A=1.27$ is the nucleon axial coupling, $m_N$ is the nucleon mass, and $f_{\pi}=92.2~$MeV is the pion decay constant.
For notation simplicity, from now on we shall drop the subscript $p$ when denoting the physical ALP and neutral pion states.
Note that while $g_{\pi^0 NN}$ in \eqref{Lphys} is modified relative to $g_{\pi^\pm NN}$ by a factor of $c_\alpha$, we are interested in the regime of $f_a\gg f_{\pi}$ and $|m_{\pi^0}-m_a|\gtrsim 10$~MeV, where\footnote{In eq.\,\eqref{mixing}, $m_{\pi^0}$ and $m_a$ denote the Lagrangian mass parameters before diagonalization. For $f_a \gg f_\pi$ and $m_{\pi^0} \neq m_a$, they provide a very good approximation to the physical ALP and neutral pion masses, and we will use them interchangeably.}:
\begin{equation}
s_{\alpha}\approx \frac{f_{\pi}}{f_a}\frac{m^2_{\pi^0}}{m^2_{\pi^0}-m^2_a},\hspace{0.5cm}
c_{\alpha}\approx 1+\mathcal{O}(s_\alpha^2). \label{mixing}
\end{equation}
Existing limits on $|g_{\pi^\pm NN}-g_{\pi^0NN}|$ (see, e.g., \cite{ParticleDataGroup:2022pth,Reinert:2020mcu}) put only a very mild constraint on the ALP decay constant of $f_a\gtrsim\mathcal{O}(10)$~GeV for $|m_{\pi^0}-m_a|\sim 10$~MeV.

The interactions in \eqref{Lphys} induce both SI and SD amplitudes for $\mu+N\to e+N$. The corresponding one-body operators are given by:
\begin{widetext}
\begin{eqnarray}
\mathcal O_{\rm SI}(q) &=&- \frac{i}{f_a^2} \frac{2}{m_\mu} \bar e \Big( C_{DL}(q) \sigma^{\mu \nu} P_R + C_{DR}(q) \sigma^{\mu \nu} P_L \Big) \mu \, q_\mu\, \bar N \gamma_\nu \left(\frac{1 + \tau^3}{2}\right) N,\label{OSI}\\
\mathcal O_{\rm SD}( q )&=& -\frac{i}{f_a^2}\, \bar e \left( C_S( q)  + C_P(q)  \gamma^5 \right) \mu \,  \frac{m_N}{m_\mu}  \bar N i \gamma^5 \tau^3 N,\label{OSD}
\end{eqnarray}
\end{widetext}
where $\mathcal O_{\rm SI}$ is induced by the CLFV dipole operator (left in Fig. \ref{dipole_4f}); $\mathcal O_{\rm SD}$ is the contribution from $t$-channel $a/\pi^0$ exchange (right in Fig. \ref{dipole_4f}); and $q = p_\mu - p_e$ is  the momentum transfer.
At $q^2=-m_\mu^2$, the couplings appearing in \eqref{OSI} are given by 
\begin{align}
C_{DL(DR)}=-\frac{\alpha_{\rm em}}{8\pi}(a_{e\mu}\pm v_{e\mu})a_{\mu\mu}\left[c^2_{\alpha}g_1(x_a)+s^2_{\alpha}g_1(x_{\pi^0}) \right], \label{pseudo}
\end{align}
with $+(-)$ for $C_{DL}$($C_{DR}$), and $x_{X}\equiv m^2_X/m^2_{\mu}$. The loop function $g_1(x)$ is given in Ref.\,\cite{Cornella:2019uxs}, and it has asymptotic limits $g_1(x)\to1$ when $x\to 0$, and $g_1(x)\to0$ when $x\to \infty$.  At the pion mass, $g(x_{\pi^0})=0.55$. On the other hand, those in \eqref{OSD} are expressed by
\begin{eqnarray}
C_{S}&=&g_As_{\alpha}c_{\alpha}\frac{f_a}{f_{\pi}}\frac{m^2_{\mu}(m^2_{\pi^0}-m^2_a)}{(m^2_{\mu}+m^2_a)(m^2_{\mu}+m^2_{\pi^0})}v_{e\mu}, \label{scalar} \\
C_{P}&=&g_As_{\alpha}c_{\alpha}\frac{f_a}{f_{\pi}}\frac{m^2_{\mu}(m^2_{\pi^0}-m^2_a)}{(m^2_{\mu}+m^2_a)(m^2_{\mu}+m^2_{\pi^0})}a_{e\mu}. \label{pseudo}
\end{eqnarray}

Before discussing the rates of $\mu\to e$ conversion in greater detail, we can easily see the dominance of the $a/\pi^0$ exchange contribution over the dipole contribution by looking at the ratio of the squared-magnitudes of their respective Wilson coefficients, 
\begin{align}
\frac{|C_{S/P}|^2}{|C_{D}|^2}=&~
t^2_{\alpha}\frac{f_a^2}{f_{\pi}^2}
\frac{g^2_A m^4_{\mu} (m^2_{\pi^0}-m^2_a)^2}{\frac{\alpha^2_{\rm em}}{64\pi^2}g^2_1(x_a)(m^2_{\mu}+m^2_a)^2(m^2_{\mu}+m^2_{\pi^0})^2}, \label{comparison1}
\end{align}
where we took $a_{\mu\mu}=v_{e\mu}=a_{e\mu}$ and neglected the sub-leading $\pi^0$-loop contribution to $C_D$. In the light ALP mass region ($m_a\ll m_{\pi^0}$) for which $t_{\alpha}\sim f_{\pi}/f_a$, we have
\begin{equation}
\frac{|C_{S/P}|^2}{|C_{D}|^2}\sim \frac{g_A^2}{\frac{\alpha^2_{\rm em}}{64\pi^2}} \sim O(10^7). \label{comparison2}
\end{equation}
This simple ratio indicates the main intuition alluded to in the introduction: while the $a/\pi^0$ exchange amplitude originates from the tree-level process with a relatively large coupling $g_A$, the dipole-mediated amplitude suffers from the QED one-loop suppression.

We now turn to the $\mu \to e $ conversion rate generated by the dipole and $a/\pi^0$ exchange contributions. Following \cite{Haxton:2022piv}, we have
\begin{eqnarray}
&&{\Gamma}\simeq \frac{m^5_{\mu}}{f^4_a}\frac{(Z_{\rm eff}\alpha_{\rm em})^3}{2\pi^2}\Bigg[W_{MM}\left(|C_{D}^{(+)}|^2+|C_{D}^{(-)}|^2 \right)\nonumber\\
%&\hspace{2.5cm}
&&~~~~~~~~~~~~~~~~~~~~+\frac{1}{4} W^{11}_{\Sigma^{\prime\prime}\Sigma^{\prime\prime}}\left(|C_{S}|^2+|C_{P}|^2 \right)\Bigg], \label{rate}
\end{eqnarray}
with $C^{(\pm)}_{D}\equiv C_{DL}\pm C_{DR}$. The first and second lines correspond to the SI and SD contributions, respectively. $W_{MM}$ and $W^{11}_{\Sigma^{\prime\prime}\Sigma^{\prime\prime}}$ are the nuclear response functions and $Z_{\rm eff}$ is an effective atomic number.\footnote{In \cite{Haxton:2022piv}, the dipole operator corresponds to $d_{9,17}$, and the pseudoscalar exchange operator to $d_{2,4}$.} The above decay rate is approximated by replacing the momentum transfer and effective electron momentum with $m_{\mu}$ in the complete expression \cite{Haxton:2022piv}. In particular, for $^{27}_{13}$Al, $Z_{\rm eff}=11.8036$, $W_{MM}=61.67$, and $W^{11}_{\Sigma^{\prime\prime}\Sigma^{\prime\prime}}=9.2\times 10^{-2}$,\footnote{We calculated response functions using the Mathematica script: \url{https://github.com/Berkeley-Electroweak-Physics/Mu2e}.} and one finds
\begin{equation}
\frac{\Gamma_{S/P}}{\Gamma_{D}}\simeq 4\times 10^{-4}\,\frac{|C_{S/P}|^2}{|C_{D}|^2}.
\end{equation}

%---------------- Figures ---------------------
\begin{figure}[t!]
\begin{center}
\includegraphics[scale=0.53]{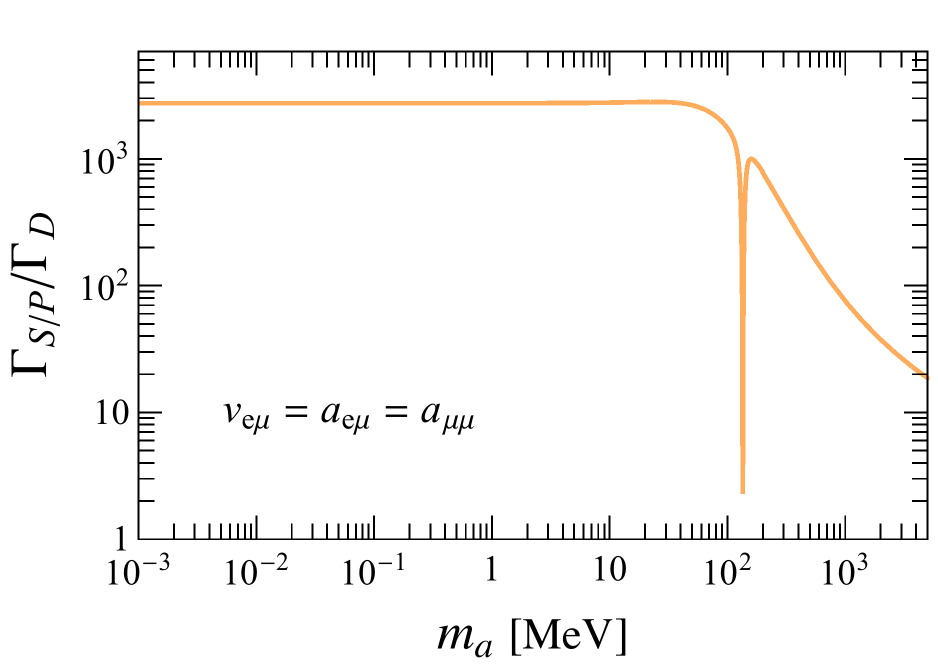}
\end{center}
\vspace{-1em}
\caption{Ratio of $\mu+N\to e+N$ transition rates, $\Gamma_{S/P}/\Gamma_D$, as a function of the ALP mass, $m_a$, assuming $^{27}_{13}$Al as the nuclear target.}
\vspace{-1em}
\label{ratio}
\end{figure}
%------------------------------------------------

Fig.\,\ref{ratio} shows the ratio ${\Gamma_{S/P}}/{\Gamma_{D}}$ in $^{27}_{13}$Al, assuming $v_{e\mu}=a_{e\mu}=a_{\mu\mu}$. For $m_a$ ranging from the ultralight regime up to $\mathcal{O}$(GeV), the ratio ranges from $O(10)$ to $O(10^3)$, clearly indicating that the $\mu\to e$ conversion process is dominated by the contribution from $a/\pi^0$ exchange in this mass range.
However, as $m_a$ becomes heavier, this contribution quickly drops off compared to the dipole one due to log dependence in the loop function.

%-----------------------------------------------------------------------------------------------------------------------------------
%	Results
%-----------------------------------------------------------------------------------------------------------------------------------
%---------------- Figures ---------------------
\begin{figure}[t!]
\begin{center}
\includegraphics[scale=0.5]{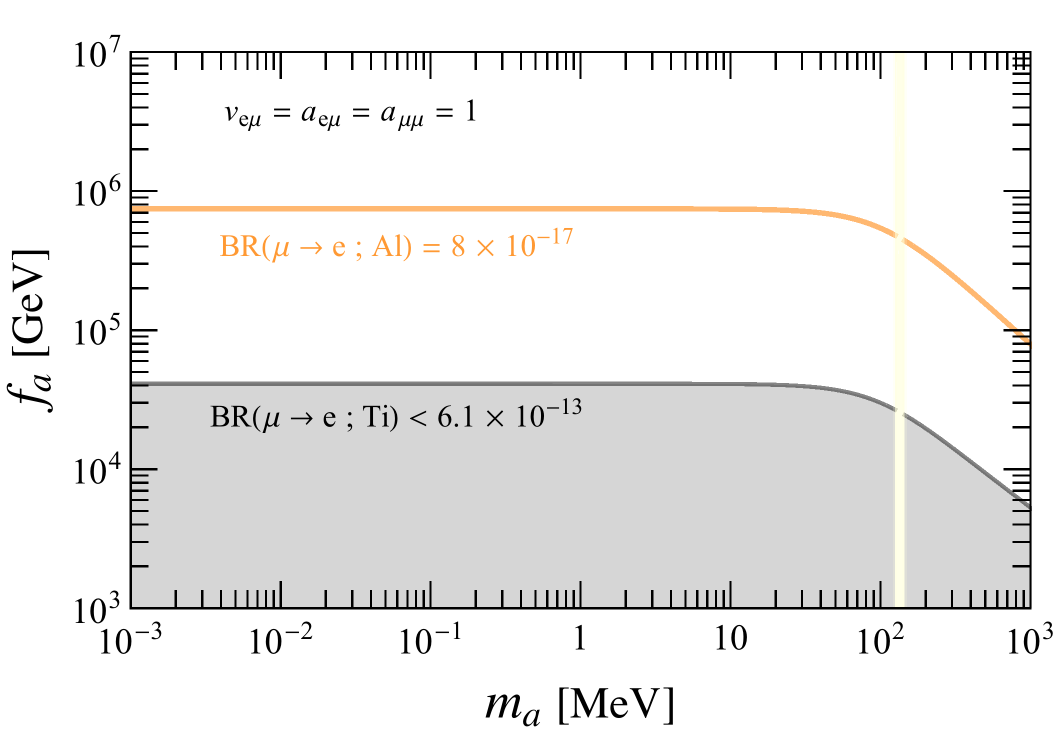}
\end{center}
\vspace{-1em}
\caption{Current (shaded region) and projected (orange line) exclusion limits from $\mu\to e$ conversion searches on the ALP parameter space of $m_a$ vs $f_a$.
The region of $|m_{\pi^0}-m_a|< 10$~MeV, for which the approximation in eq.\,(\ref{mixing}) is not valid, has been masked out. }
\label{limitfa}
\end{figure}
%------------------------------------------------

\vspace{1em}
\noindent\emph{Results---}
The branching ratio for $\mu\to e$ conversion normalized to the muon capture rate is given by
\begin{equation}
{\rm BR}(\mu\to e)=\frac{\Gamma_{\rm conv}(\mu + (A,Z) \to e+ (A,Z))}{\Gamma_{\rm capt}(\mu + (A,Z) \to \nu_{\mu} (A,Z-1))}.
\end{equation}
The capture rate is experimentally determined to be \cite{Suzuki:1987jf}
\begin{eqnarray}
%\Gamma_{\rm capt}(\text{Au})&=&8.60\times 10^{-15}~\text{MeV},\\
\Gamma_{\rm capt}(\text{Ti})&=&1.71\times 10^{-15}~\text{MeV},\\
\Gamma_{\rm capt}(\text{Al})&=&4.63\times 10^{-16}~\text{MeV}.
\end{eqnarray}
The current experimental bound for Ti is \cite{Wintz:1998rp}
\begin{eqnarray}
{\rm BR}(\mu\to e;~{\rm Ti})&<&6.1\times 10^{-13}.
\end{eqnarray}

In Fig.\,\ref{limitfa} we show existing constraints from $\mu\to e$ conversion in titanium (fixing $v_{e\mu}=a_{e\mu}=a_{\mu\mu}=1$) in the parameter space of ALP mass $m_a$ vs ALP decay constant $f_a$. We also present the future reach of $\mu\to e$ conversion in aluminum assuming a projected sensitivity of future measurements of $\text{BR}(\mu\to e;~{\rm Al})=8.0\times 10^{-17}$.

\begin{figure}[t!]
\begin{center}
\includegraphics[scale=0.4]{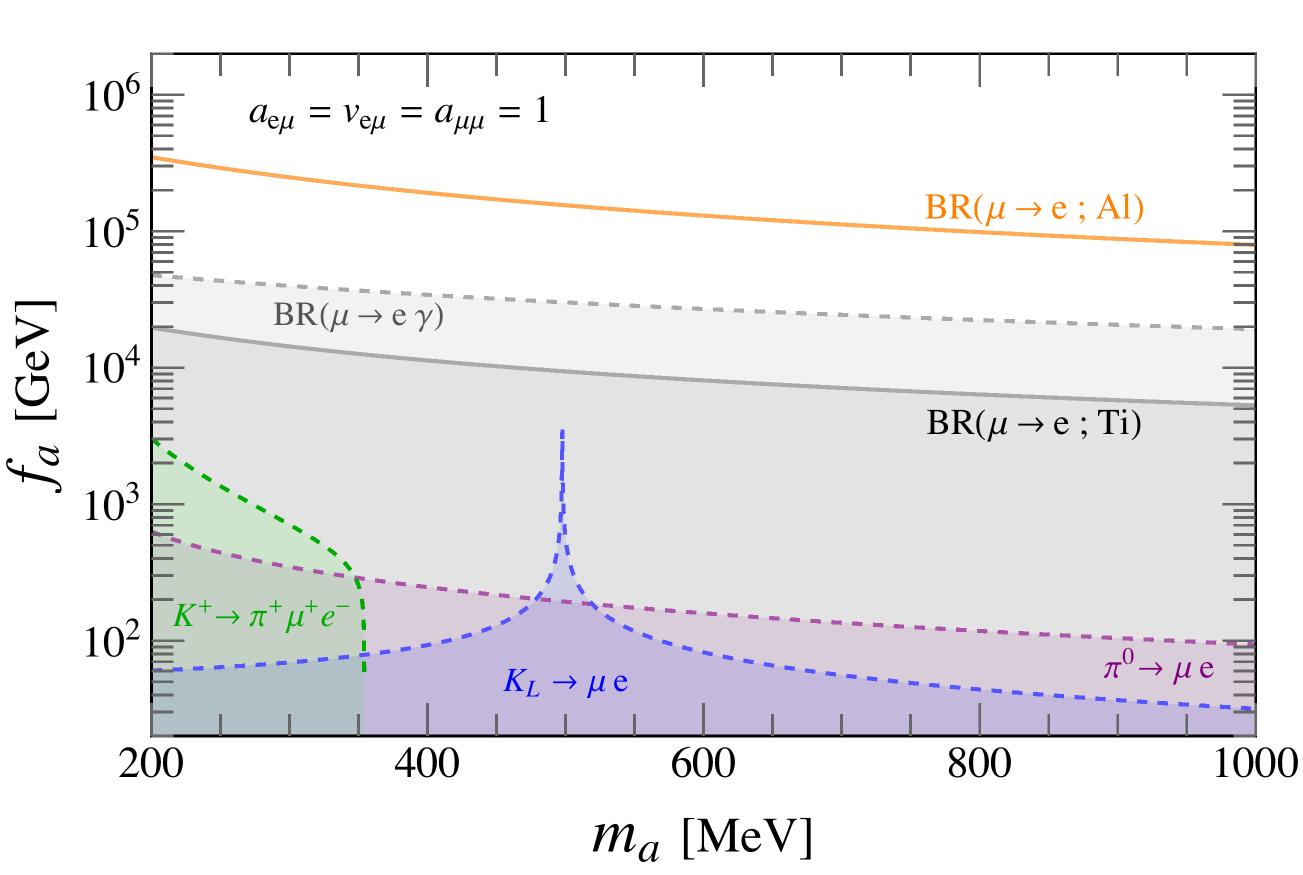}
\end{center}
\vspace{-1em}
\caption{Similar to Fig.\,\ref{limitfa}, but focusing on the region of $m_a>200$~MeV, and showing additional limits from searches for rare CLFV decays (shaded regions delimited by dashed lines): $\mu\to e \gamma$ (light gray), $K^+\to \pi^+ \mu^+ e^-$ (green), $K_L\to \mu e$ (blue), and $\pi^0\to \mu e$ (purple). Weaker bounds from LFC Kaon decays, such as $K^+\to \pi^+ \mu\mu$, $K_L\to \mu \mu$, are not shown.}
\label{limitfa2}
\end{figure}

Several other constraints in the ALP parameter space can be competitive with or dominant relative to the $\mu+N\to e+N$ reach. In particular, for $m_a<m_\mu- m_e$, existing constraints from $\mu\to e a$ decays are quite severe, excluding $f_a\gtrsim\mathcal{O}(10^9)$~GeV \cite{Calibbi:2020jvd, Perrevoort:2018okj}. Furthermore, future searches for $\mu\to e a$ are projected to reach $f_a \sim \mathcal{O}(10^{10})$~GeV, far beyond any future reach of $\mu\to e$ conversion searches.  Hence, in Fig.\,\ref{limitfa2}, we focus on the region of $m_a=(200-1000)$~MeV, where the projected sensitivity of $\mu+\text{Al}\to e+\text{Al}$ searches will be able to probe unexplored parameter space of ALP models with CLFV and hadronic couplings.

In particular, in Fig.\,\ref{limitfa2} we show additional exclusions in our ALP model from bounds on $\mu\to e\gamma$ and LFV meson decays. The branching ratio for $\mu\to e\gamma$ has a current upper bound of ${\rm BR}(\mu^+\to e^+\gamma)<4.2\times 10^{-13}$ \cite{MEG:2016leq}, which translates into a marginally stronger limit on $f_a$ compared to current limits from $\mu+N\to e+N$. However, the future reach of $\mu\to e\gamma$ searches on $f_a$ are expected to improve only mildly by a factor of $\sim$~2, \cite{MEGII:2018kmf, MEGII:2021fah}, whereas upcoming $\mu+N\to e+N$ searches will improve the reach on $f_a$ by over an order of magnitude.

In the meson sector, $a-\pi^0$ mixing induces rare LFV meson decays such as $K^+\to\pi^+(a\to\mu e)$, $K_L\to \mu e$ and $\pi^0\to\mu e$, whose current experimental upper bounds are given by \cite{KTeV:2007cvy,BNL:1998apv,Sher:2005sp}:
\begin{subequations}
\begin{align}
\text{BR}(\pi^0\to \mu e)&<3.6\times 10^{-10},\label{BRKpimue}\\
\text{BR}(K_{\!L}\to \mu e)&<4.7\times 10^{-12},\label{BRKLmue}\\
\text{BR}(K^+\to \pi^+ \mu^+ e^-)&<1.3\times 10^{-11}.\label{pimue}
\end{align}
\end{subequations}
Using the $\chi$PT method in \cite{Alves:2020xhf}, the upper bounds listed in (\ref{BRKpimue}–\ref{pimue}) can be straightforwardly translated into limits on $m_a$ vs $f_a$, and are shown in Fig.\,\ref{limitfa2}. 

Other bounds not shown in Fig.\,\ref{limitfa2} come from rare lepton flavor-conserving (LFC) Kaon decays such as $K^+\to \pi^+ (a\to\mu\mu)$ and $K_{\!L}\to \mu\mu$. These processes are induced by the LFC ALP coupling to muons in \eqref{diagonal_ALP} combined with $a-\pi^0$ mixing. Constraints on $f_a$ from these LFC decays are weaker than their LFV counterparts by about an order of magnitude \cite{NA482:2016sfh,E871:2000wvm}, and therefore we omit their exclusions from Fig. \ref{limitfa2}.

%---------------- Figures ---------------------
%------------------------------------------------

While the $f_a$ constraints in Figs.~\ref{limitfa} and \ref{limitfa2} assumed $a_{e\mu}=v_{e\mu}=1$, they can be easily recast as limits on $a_{e\mu}$ and $v_{e\mu}$ at fixed $f_a$ by noting that the $\mu\to e$ conversion rate scales as $\Gamma(\mu N\to eN)\propto (|a_{e\mu}|^2+|v_{e\mu}|^2)/f_a^4$. For instance, fixing $f_a=1$~TeV, the upper limits are $|v_{e\mu}|,~|a_{e\mu}|\lesssim \mathcal{O}(10^{-2}-10^{-3})$ for $m_a=(200-1000)$~MeV, and are expected to improve by two orders of magnitude with future $\mu\to e$ conversion searches.

Finally, as discussed in \cite{Davidson:2017nrp, Haxton:2022piv}, the use of different nuclear targets in $\mu\to e$ conversion searches could help to disentangle the underlying ALP interactions. In particular, a finite nuclear spin is sensitive to both SI and SD interactions. The Mu2e and COMET experiments will use an aluminum target, which has a 100\% natural abundance of the isotope {$^{27\!}$Al} with a nuclear spin $J=5/2$. On the other hand, natural titanium is predominantly composed of the {$^{48}$Ti} isotope ($\sim$74\% natural abundance), which has $J=0$. Hence, the SD contribution to $\mu\to e$ conversion in a titanium target is relatively suppressed compared to $^{27}$Al. Targets with $J=0$ nuclear isotopes, such as O and Ca,  would be suitable to isolate the dipole contribution.

%-----------------------------------------------------------------------------------------------------------------------------------
%	Conclusions
%-----------------------------------------------------------------------------------------------------------------------------------
\paragraph*{Conclusions---} We have studied $\mu\to e$ conversion process arising from one-loop dipole and tree-level interactions induced by CLFV ALP interactions. While the dipole operator originates from purely leptonic ALP interactions, $a/\pi^0$ exchange interactions can also exist if ALP interacts with the SM light quarks. Assuming ALP couplings are $O(1)$, we find that the tree-level interaction dominates the conversion process over the dipole one due to the QED one-loop suppression factor. The current experimental limit indicates $f_a\gtrsim O(1)~$TeV for $m_a>m_{\mu}$, which is relatively strong compared to other existing bounds from CLFV meson decays. The next-generation searches are expected to reach $f_a\sim O(100)~$TeV in the heavy $m_a$ region, which exhibits the highest potential to probe CLFV ALP interactions.
Although the current study considered an ALP model with exclusively isovector couplings to light quarks, we expect that our qualitative conclusions would hold even if heavy quark and/or gluon couplings are also present. 

\vspace{1em}
\noindent{\it Acknowledgments---} We thank Daniele Alves for helpful discussions and input on the manuscript. We are also grateful to Evan Rule for valuable discussions, especially on numerical calculations of response functions using the Mathematica script: \url{https://github.com/Berkeley-Electroweak-Physics/Mu2e}. This work was supported by the US Department of Energy Office of Science Nuclear Physics, an Early Career LDRD Award, and the LDRD program at Los Alamos National Laboratory. Los Alamos National Laboratory is operated by Triad National Security, LLC, for the National Nuclear Security Administration of the U.S. Department of Energy (Contract No. 89233218CNA000001).

\bibliography{refs}

\end{document}